\documentclass[aip,jap,reprint]{revtex4-1}
\usepackage{graphicx}
\usepackage{soul,color}
\sethlcolor{cyan}

\begin{document}

\title{Ferroelectricity and piezoelectricity in monolayers and nanoplatelets of SnS}

\author{Alexander I. Lebedev}
\email[]{swan@scon155.phys.msu.ru}
\affiliation{Physics Department, Moscow State University, 119991 Moscow, Russia}

\date{\today}

\begin{abstract}
The ground-state structure of monolayers and nanoplatelets of SnS with a thickness
from two to five monolayers is calculated from first principles. It is shown that
nanoobjects with only odd number of monolayers are ferroelectric. The ferroelectric,
piezoelectric, and elastic properties of these polar structures are calculated.
The appearance of polarization in these nanoobjects is explained by an uncompensated
polarization that exists in an antiferroelectric structure of bulk SnS. The mechanism
of ferroelectricity, in which the ferroelectric distortion is associated with
short-range ordering of lone pairs, can be regarded as a way of creating
ferroelectrics with high Curie temperature.
\end{abstract}


\maketitle

\section{Introduction}

In recent years, quasi-two-dimensional (2D) structures---monolayers and
nanoplatelets with a thickness of a few monolayers---have gained a considerable
attention.~\cite{JAmChemSoc.130.16504,JMaterChem.19.2503,
CritRevSolidStateMaterSci.35.52,ChemSocRev.42.1934}  These structures have
demonstrated many interesting properties which differ much from those of bulk
crystals. Tin sulfide SnS, which in the bulk form is a semiconductor with a
layered structure, can be split into monolayers and nanoplatelets by chemical
methods.~\cite{JAmChemSoc.135.11634,JAmChemSoc.137.9943,JAmChemSoc.137.12689}
The interest to SnS quasi-2D structures is stimulated by their thermoelectric,
electric, and optical properties~\cite{JApplPhys.113.233507,ApplPhysLett.105.042103,
PhysRevB.92.085406,JChemPhys.144.114708,JPhysChemC.120.18841,NanoLett.16.3236,
JApplPhys.121.034302,PhysRevB.95.235434,ACSNano.11.2219,PhysRevLett.118.227401}
and the predicted ferroelectricity in SnS monolayers.~\cite{ApplPhysLett.107.173104,
PhysRevB.94.035304}  The latter suggests a possible application of piezoelectric
properties of these structures. As concerns to SnS nanoplatelets containing more
than one monolayer, only the optical properties of these nanoobjects have been
studied.~\cite{JApplPhys.113.233507,PhysRevB.92.085406,JAmChemSoc.137.12689}

At room temperature, SnS crystallizes in the $Pnma$ structure. At 878~K, it
undergoes a second-order structural phase transition into a high-temperature
$Cmcm$ phase.~\cite{JPhysChemSolids.47.879}  This phase has a distorted NaCl
structure and consists of two-layer slabs of this structure which are shifted
against the adjacent slabs by a half of the translation vector of the NaCl
structure [Fig.~\ref{fig1}(a)]. The chemical bonding between the slabs is
believed to be of van der Waals type; such structures can be easily split into
the slabs (monolayers). The $Fm{\bar 3}m$ structure of SnS is another metastable
phase which was obtained in epitaxial films.~\cite{ApplPhysLett.10.282}

The use of GGA-PBE PAW pseudopotentials in earlier first-principles calculations
of SnS nanostructures~\cite{JApplPhys.121.034302,JChemPhys.144.114708,
NanoLett.16.3236,ApplPhysLett.107.173104,PhysRevLett.117.097601}
resulted in the unit cell volume of bulk $Pnma$ phase, which strongly (by
$\sim$6\%) exceeded the experimental one. This can lead to an overestimation of
the ferroelectric instability and other ferroelectric properties. Our tests
have shown that the local density approximation (LDA) PAW pseudopotentials
underestimate the unit cell volume
by 6.2\% and the energies of the ferroelectric ordering in SnS monolayers
calculated using the GGA and LDA functionals differ by more than 40~times.%
    \footnote{Our calculations using the GGA-PBE PAW pseudopotentials taken from
    Ref.~\onlinecite{ComputMaterSci.81.446} gave the energy gain of 20.4~meV per
    unit cell for SnS monolayer with the $Pmn2_1$ structure. Using of the LDA PAW
    pseudopotentials from the same source gave much lower energy, 0.48~meV
    per unit cell.}
On the other hand, the applicability of the GGA functional to two-dimensional
systems is questionable because an inaccuracy in its correlation part. For these
systems, the LDA approximation usually gives better results.~\cite{PhysRevLett.96.136404}
In this work, we used norm-conserving LDA pseudopotentials, for which the unit
cell volume of bulk SnS is underestimated by only
$\sim$1\% against the experimental data extrapolated to $T = 0$.
In addition to the studies of ferroelectric and piezoelectric properties of
SnS monolayers, in this work the ground-state structure and properties of SnS
nanoplatelets with a thickness from 2 to 5~monolayers (MLs) are calculated and
the origin of ``intermittent'' ferroelectricity revealed in these nanoplatelets
is discussed.

\section{Calculation technique}

The first-principles calculations presented in this work were carried out within
the density functional theory using the \texttt{ABINIT} software package and
norm-conserving pseudopotentials constructed using the RKKJ scheme~\cite{PhysRevB.41.1227}
in the local density approximation.~\cite{PhysSolidState.51.362} The
cutoff energy was 30~Ha (816~eV) and the integration over the Brillouin zone
was performed using an 8$\times$8$\times$4 Monkhorst--Pack mesh. Equilibrium
lattice parameters and atomic positions were obtained by relaxing the forces
acting on the atoms to a value below $2 \cdot 10^{-6}$~Ha/Bohr (0.1~meV/{\AA}).
The phonon spectra and the elastic and piezoelectric moduli were calculated
using the density-functional perturbation theory. The Berry phase method was
used to calculate the polarization.

\section{Results}
\subsection{Phase relations for bulk SnS}

The calculation of the energies of different possible structures of bulk SnS
shows that the $Pnma$ structure is the ground-state structure, in agreement with
experiment. The energies of the $Cmcm$ and $Fm{\bar 3}m$ phases are by 20 and
49~meV per formula unit higher than that of the $Pnma$ phase. The calculation
of the phonon spectra and elastic tensors for these phases shows that the
$Fm{\bar 3}m$ and $Pnma$ phases fulfill the stability criterion, i.e., all optical
phonon frequencies in the Brillouin zone in them are positive and their structures
are mechanically stable. This suggests that the $Fm{\bar 3}m$ phase which has a
higher energy is metastable. As concerns to the $Cmcm$ phase, two instabilities
are observed in its phonon spectrum: the ferroelectric one at the center of the
Brillouin zone (the phonon frequency is 63$i$~cm$^{-1}$) and another one at the
$Y$~point on its boundary (the phonon frequency is 35$i$~cm$^{-1}$). The
ferroelectric instability results in a phase transition to a metastable $Cmc2_1$
phase whose energy is by 15.5~meV higher than that of the $Pnma$ phase. The
condensation of the unstable phonon at the $Y$~point transforms the structure
into the ground-state $Pnma$ phase. These two instabilities in the $Cmcm$ phase
compete with each other; the appearance of structural distortions described by
the unstable phonon at the $Y$~point suppresses the ferroelectric instability.%
    \footnote{Large Born effective charge of Sn in the $Cmcm$ phase
    ($Z^* = {}$5.4--5.7 in the $xy$ plane) favors the appearance of ferroelectricity,
    but the strong distortion of the structure associated with the stereochemical
    activity of the $s^2$ lone pair of Sn$^{2+}$ ions makes this compound
    antiferroelectric.}

\begin{figure}
\includegraphics{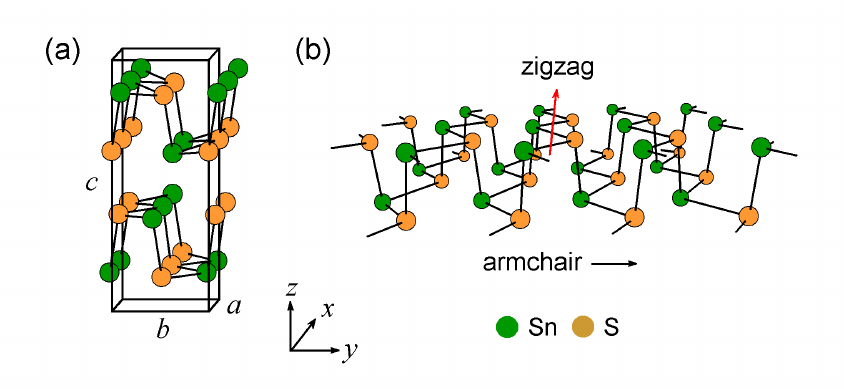}
\caption{Structures of (a) bulk SnS and (b) SnS monolayer.}
\label{fig1}
\end{figure}

A~tendency of the SnS structure to be split into separate monolayers when
stretching it along the $z$~axis is illustrated by a divergence of the
$S_{33}$ component of the elastic compliance tensor of the $Pnma$ phase
at a certain critical pressure $p_c$. This pressure can be determined as a
maximum internal pressure in the unit cell, which can be obtained by changing
the $c$~lattice parameter when relaxing atomic positions and two other lattice
parameters. The calculations show that $p_c = 28.2$~kbar for the $Pnma$ phase.
The splitting of the metastable $Fm{\bar 3}m$ phase is a more complicated
process because the stretched sample (whose symmetry decreases to $I4/mmm$ under
strain) exhibits a ferroelectric phase transition at 10.1~kbar and then a
first-order phase transition to the $P4/nmm$ phase. The $p_c$ value for this
phase is 25~kbar.

\subsection{Phase relations for SnS monolayers}

\begin{figure}
\includegraphics{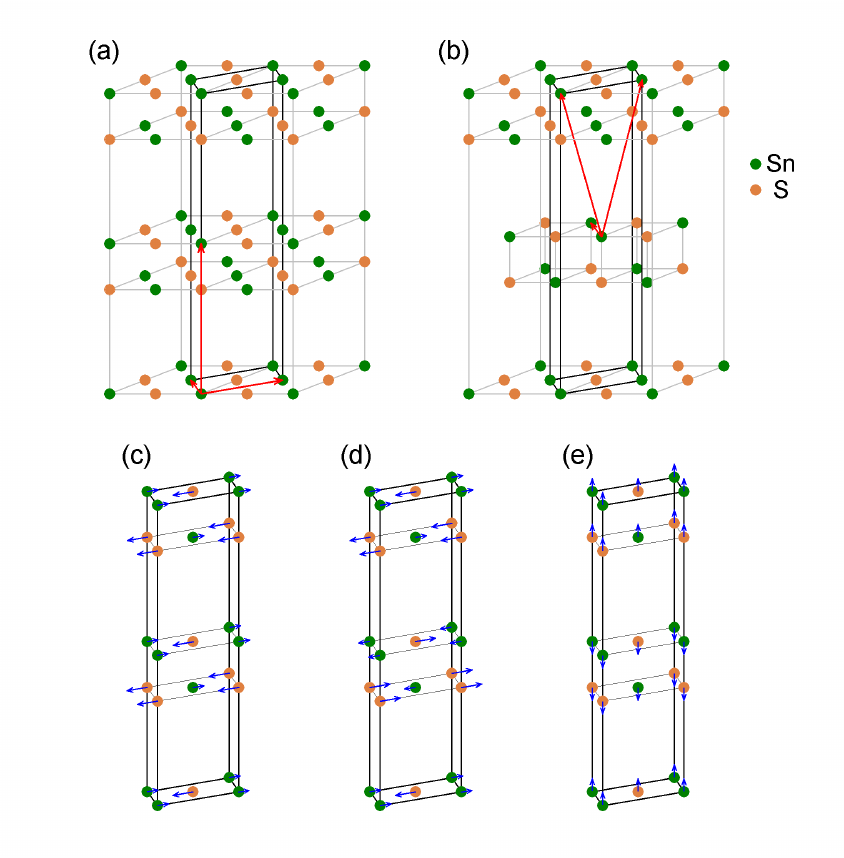}
\caption{[(a), (b)] The supercells used to study the properties of SnS monolayers.
The translation vectors of the (a) $P4/nmm$ and (b) $Cmcm$ primitive units cells are
shown by red lines. The eigenvectors of (c) $E_u$, (d) $Z_5^-$, and (e) $Z_1^-$
unstable modes found in the $P4/nmm$ structure.}
\label{fig2}
\end{figure}

To determine the ground-state structure of SnS monolayers and calculate their
physical properties, the supercells obtained by strong stretching of the
$Fm{\bar 3}m$, $Cmcm$, and $Pnma$ structures of bulk SnS along the $z$~axis
were considered. As the \texttt{ABINIT} program requires the optimizable
translation vectors to be orthogonal to the translation vector that is fixed,
supercells containing four formula units were used for $Fm{\bar 3}m$ and
$Cmcm$ structures [Figs.~\ref{fig2}(a) and 2(b)].
The stretching of the supercells results in appearance of two monolayers
separated by vacuum gaps. After splitting, the symmetry of the $Fm{\bar 3}m$
supercell is reduced to $P4/nmm$,
whereas the $Cmcm$ and $Pnma$ supercells retain their symmetry. When increasing
the $c$ lattice parameter, the difference between the $a$ and $b$ lattice
parameters in the $Cmcm$ supercell becomes very small ($<$0.0001~{\AA} at
$c = 26.46$~{\AA}). The comparison of the evolution of these supercells will
help us later to reveal different properties of nanoplatelets prepared from
monolayers conjugated at the interface by different ways.

\begin{table}
\caption{\label{table1}Energies of different phases of SnS monolayers (in meV
per primitive unit cell). The energy of the $P4/nmm$ phase is taken as the
energy origin.}
\begin{ruledtabular}
\begin{tabular}{cccccc}
Phase     & Displacement &\multicolumn{4}{c}{$c$ ({\AA})} \\
\cline{3-6}
          & direction & 18.52 & 21.17     & 26.46     & 31.75 \\
\hline
$P4/nmm$  & --- &  0      &    0      &    0      &    0 \\
$Cmcm$    & ---   & $-$0.204  &   +0.002  &   +0.001  &   +0.001 \\
$P2_1/c$  & [110] & $-$0.984\footnotemark[1] & $-$0.766\footnotemark[1] & $-$0.752 & $-$0.757 \\
$Cmca$    & [110] & $-$0.759  & $-$0.771  & $-$0.753  & $-$0.756 \\
$Abm2$    & [110] & $-$0.787  & $-$0.786  & $-$0.760  & $-$0.743 \\
$Cc$      & [110] & $-$0.968  & $-$0.797  & $-$0.765  & $-$0.766 \\
$Pmn2_1$  & [100],[010] & $-$1.361 & $-$1.336 & $-$1.320 & $-$1.304 \\
$Ama2$    & [010] & $-$1.539  & $-$1.351  & $-$1.320  & $-$1.310 \\
$Cmc2_1$  & [100] & $-$1.540  & $-$1.350  & $-$1.320  & $-$1.310 \\
$Pnnm$    & [100] & $-$1.521  & $-$1.347  & $-$1.323  & $-$1.312 \\
$Pnma$    & [010] & $-$1.559  & $-$1.347  & $-$1.322  & $-$1.314 \\
\end{tabular}
\end{ruledtabular}
\footnotetext[1]{This phase is unstable against its transformation into
the $Pnma$ or $Pnnm$ phase in supercells with small vacuum gaps. At large
vacuum gaps, it becomes more stable. The reported energies were calculated for
the case of $P_x \approx P_y$ (the condition that is satisfied at large vacuum
gaps).}
\end{table}

An increase in the thickness of the vacuum gap changes the energies of the
discussed supercells. For supercells containing two monolayers, these changes
saturate above $c = 25.4$~{\AA}. This suggests
that the thickness of the vacuum gap of $\sim$9.5~{\AA} is sufficient to neglect
the interaction between monolayers. When increasing~$c$, the energy difference
between the $P4/nmm$ and $Cmcm$ phases abruptly decreases, but the energy of
the $Pnma$ phase remains lower (Table~\ref{table1}). The weakness of the
interaction between monolayers is confirmed by the low residual pressure in the
supercells. At $c = 26.46$~{\AA} it is 0.04~kbar, which is about three orders
of magnitude lower than the critical pressure $p_c$.

\begin{figure}
\includegraphics{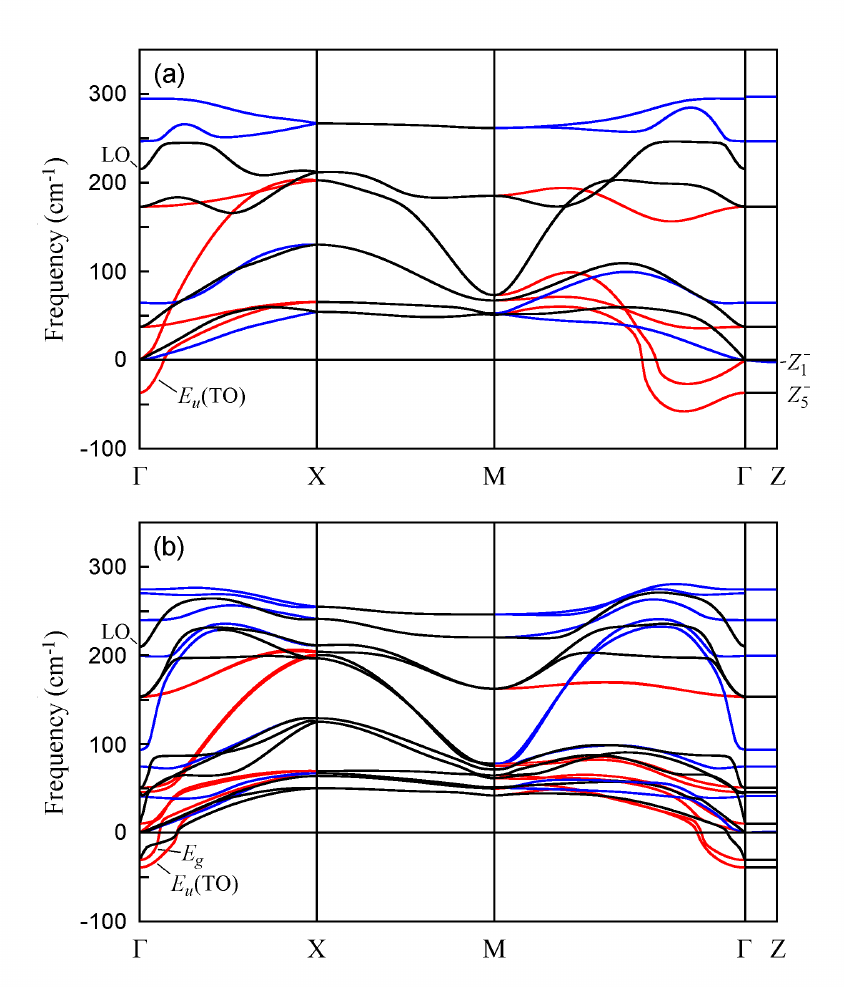}
\caption{Phonon spectra of nanoplatelets with a thickness of (a) 1ML and (b) 2ML
and the $P4/nmm$ structure. Calculations were performed on supercells with
the $c$ lattice parameter of 15.88~{\AA} and 21.17~{\AA}, respectively.
Because of a strong intermixing of longitudinal and out-of-plane atomic
displacements, the polarization of modes is well defined only near the
$\Gamma$~point. The red lines are transverse modes with in-plane polarization,
the blue lines are transverse modes with out-of-plane polarization, and the
black lines are longitudinal modes.}
\label{fig3}
\end{figure}

The calculations of the phonon spectra were performed using primitive unit
cells extracted from the supercells. The calculations for a SnS monolayer
modeled with the $P4/nmm$ supercell find a doubly degenerate unstable
ferroelectric $E_u$ mode with a frequency of 37$i$~cm$^{-1}$ at the $\Gamma$~point
and two unstable modes at the $Z$~point of the Brillouin zone [Fig.~\ref{fig3}(a)]:%
    \footnote{In order to get better accuracy for phonon frequencies on the
    boundary of the Brillouin zone, the response function calculations were
    also performed on supercells containing two monolayers in the unit cell. In this
    case, the $Z$~point of the Brillouin zone is projected on the $\Gamma$~point.}
a doubly degenerate antiferroelectric $Z_5^-$ mode and a weak nondegenerate
$Z_1^-$ mode with a frequency of 0.199$i$~cm$^{-1}$ describing an acoustic-like
antiparallel shift of adjacent monolayers along the $z$~axis. The eigenvectors
of these modes are shown in Figs.~\ref{fig2}(c)--2(e). The latter mode
describes the coalescence of monolayers into a nanoplatelet, and this is why it
will be called hereafter the coalescence mode. In several cases, a very weak
($\sim$0.03$i$~cm$^{-1}$) unstable doubly degenerate mode describing an acoustic-like
antiparallel sliding of adjacent monolayers in the $xy$~plane appeared in the
spectrum. The frequency of the antiferroelectric $Z_5^-$ mode, in which the polar
atomic displacements in the $xy$~plane in adjacent monolayers are antiparallel
to each other, differs from the frequency of the ferroelectric $E_u$ mode by less
than 0.001~cm$^{-1}$. This indicates a weak interaction between polarizations
in adjacent monolayers.

The comparison of the energies of two possible $Abm2$ and $Pmn2_1$ polar phases,
into which the $P4/nmm$ structure can transform upon the ferroelectric distortion,
shows that the energy of the $Pmn2_1$ phase is lower (Table~\ref{table1}). An
analysis of the phonon spectra and elastic tensors of these two phases shows that
the $Abm2$ phase is unstable, whereas the $Pmn2_1$ phase meets the stability
criterion except for weak unstable modes at the $Z$~point resulting from a weak
interaction between the monolayer and its images. The condensation of the unstable
antiferroelectric $Z_5^-$ mode can result in appearance of the $Cmca$ and $Pnma$
phases. The $Cmca$ phase is unstable and relaxes to the $Pnma$ phase or to a
metastable $Aba2$ phase, in which the polarization in adjacent monolayers is
rotated by $\sim$90$^\circ$.

The calculations of the phonon spectrum of a SnS monolayer modeled with the
$Cmcm$ supercell find two unstable ferroelectric modes, $B_{2u}$ and $B_{3u}$,
with close energies at the $\Gamma$~point (37$i$~cm$^{-1}$) and at least three
unstable modes at the $Y$~point. These are two antiferroelectric modes whose
frequencies are very close to those of the ferroelectric modes and a weak
coalescence mode (0.197$i$~cm$^{-1}$).
In several cases, one or two very weak ($\sim$0.03$i$~cm$^{-1}$) unstable modes
describing an antiparallel sliding of adjacent monolayers in the $xy$~plane
appeared in the spectrum. An analysis of properties of three possible $Cmc2_1$,
$Ama2$, and $Cc$ polar phases, into which the $Cmcm$ structure can transform
upon the ferroelectric distortion, shows that the $Cc$ phase is unstable and
the $Cmc2_1$ and $Ama2$ phases have the same lowest energy (Table~\ref{table1}).
The coincidence in their energies is a result of nearly equal $a$ and $b$ lattice
parameters in the $Cmcm$ supercell. The condensation of unstable antiferroelectric
modes can result in appearance of the $Pnnm$, $P2_1/c$, and $Pnma$ phases. The
$P2_1/c$ phase is unstable, and the $Pnnm$ and $Pnma$ phases have close lowest
energies.

The comparison of the energies of all found phases (Table~\ref{table1}) shows
that among possible ground-state structures of SnS monolayers, the phases
with the [100] or [010] atomic displacements
are the most energetically favorable. The weak dependence on the direction
results from close $a$ and $b$ lattice parameters in the parent structures.
However, the result why the energy is nearly independent on whether the
supercell structure is ferroelectric or antiferroelectric needs explanation.

This result is not surprising since in the $c \to \infty$ limit all found
solutions are physically equivalent and differ only by the orientation of
polarization in weakly interacting adjacent monolayers. The appearance of the
non-polar $Pnma$ and $Pnnm$ phases does not contradict the polar $Cmc2_1$,
$Ama2$, and $Pmn2_1$ solutions because in the $Pnma$ and $Pnnm$ structures
there are glide planes reversing the $x$ or $y$ direction of polarization upon
a half-a-period shift along the $z$~axis. This means that the $Pnma$ and $Pnnm$
structures are actually \emph{antiferroelectric}. Taking into account that the
$Ama2$ and $Cmc2_1$ phases differ only by the polarization direction and the
$Cmc2_1$ and $Pmn2_1$ phases have the same polarization direction and differ
by the way of conjugation of adjacent monolayers, in the $c \to \infty$ limit
these differences become negligible, and the $Pmn2_1$ phase can be considered as
the ground-state structure of SnS monolayers. The polarization in this structure
is directed along the armchair direction [Fig.~\ref{fig1}(b)]. Our conclusion
agrees with the results of earlier
calculations,~\cite{ApplPhysLett.105.042103,JApplPhys.121.034302} but disagrees
with the $Pmcn$ space group incorrectly identified in
Ref.~\onlinecite{NanoLett.16.3236} (this group is non-polar).

\subsection{Coalescence of monolayers and properties of SnS nanoplatelets}
\label{sec3c}

As noted above, in the phonon spectra of all studied SnS supercells there is
a weak unstable mode whose eigenvector describes the coalescence of the
monolayers into a two-layer (2ML) nanoplatelet. The typical frequency of this
mode is 0.2--2$i$~cm$^{-1}$ for both $P4/nmm$ and $Cmcm$ supercells.

\begin{table}
\caption{\label{table2}Energies of different phases of 2ML SnS nanoplatelets
obtained from the $P4/nmm$ and $Pmma$ parent structures. The energy of two
isolated monolayers is taken as the energy origin. The energy of the most
stable phase is in boldface.}
\begin{ruledtabular}
\begin{tabular}{cccc}
Phase & Unstable phonon     & Displacement & Energy \\
      & and order parameter & direction    & (meV/2ML) \\
\hline
$P4/nmm$ & ---              & ---          & $-$295.8 \\
$Abm2$   & $E_u(\eta,\eta)$ & [110]        & $-$297.3 \\
$Pmn2_1$ & $E_u(\eta,0)$    & [100]        & $-$298.6 \\
$C2/m$   & $E_g(\eta,\eta)$ & [110]        & $-$371.8 \\
$P2_1/m$ & $E_g(\eta,0)$    & [100]        & $-$483.6 \\
\hline
$Pmma$   & ---              & ---          & $-$455.1 \\
$Pma2$   & $B_{2u}$         & [010]        & $-$455.2 \\
$Pmc2_1$ & $B_{3u}$         & [100]        & $-$464.7 \\
$P2_1/m$ & $B_{2g}$         & [010]        & {\bf $-$483.7} \\
\end{tabular}
\end{ruledtabular}
\end{table}

The coalescence of SnS monolayers into the 2ML nanoplatelet is indeed energetically
favorable. The energy gain, however, depends on how the monolayers are conjugated
at the interface. For nanoplatelets with the $Pmma$ symmetry, which are
obtained from the $Cmcm$ supercell, the energy gain is notably higher than for
nanoplatelets with the $P4/nmm$ symmetry obtained from the $P4/nmm$ supercell
(Table~\ref{table2}). We note that both energy gains are too high to explain
the chemical bonding between monolayers by van der Waals (vdW) interaction
only: typical bonding energies for vdW systems is one order of magnitude
smaller.~\cite{PhysRevLett.92.246401}  Large energy gains mean that the
nanoplatelet/monolayer surfaces need a coating to protect them against the
aggregation.

The phonon spectrum of a SnS 2ML nanoplatelet with the $P4/nmm$ symmetry
exhibits two doubly degenerate unstable modes at the $\Gamma$~point
[Fig.~\ref{fig3}(b)]: the ferroelectric $E_u$ mode (39$i$~cm$^{-1}$)
and the antiferroelectric $E_g$ mode (31$i$~cm$^{-1}$), which differ by the
orientation of polarization in adjacent monolayers.

The phonon spectrum of a SnS 2ML nanoplatelet with the $Pmma$ symmetry
exhibits three unstable modes at the $\Gamma$~point: a strong ferroelectric
$B_{3u}$ mode (52$i$~cm$^{-1}$) polarized along the zigzag direction
(with transition to the $Pmc2_1$ phase), a weak ferroelectric
$B_{2u}$ mode (8$i$~cm$^{-1}$) polarized along the armchair direction
(with transition to the $Pma2$ phase), and a $B_{2g}$ mode
(31$i$~cm$^{-1}$) describing strong antiferroelectric atomic displacements
along the armchair direction (with transition to the $P2_1/m$ phase). The
energy gain resulting from the latter distortion (28.6~meV per unit cell) is
much higher than those for both ferroelectric distortions.

As follows from Table~\ref{table2}, among the phases resulting from
condensation of the unstable modes, the $P2_1/m$ ones have the lowest energy.
The energy of the phase obtained from the $P4/nmm$ parent structure is slightly
higher than the energy of the same phase obtained from the $Pmma$ parent
structure. Taking into account that these phases are nearly the same and
the difference in their energies results from different interaction between
the nanoplatelet and its images, one can conclude that the ground-state
structure of the 2ML nanoplatelet is $P2_1/m$, and so these nanoplatelets are
not ferroelectric.%
    \footnote{The obtained results are independent of the exchange-correlation
    functional used in the calculations. The same sequence of the energies was
    obtained when modeling different low-symmetry phases of 2ML nanoplatelets
    using the GGA-PBE and dispersion-corrected GGA-PBE+D2 functionals.}
The structures of the ground state of nanoplatelets with different thickness
are given in the supplementary material.

\begin{table}
\caption{\label{table3}Energies of different phases of 3ML SnS nanoplatelets
obtained from the $P4/nmm$ and $Pmmn$ parent structures. The energy of three
isolated monolayers is taken as the energy origin. The energy of the most
stable phase is in boldface.}
\begin{ruledtabular}
\begin{tabular}{cccc}
Phase & Unstable phonon     & Displacement & Energy \\
      & and order parameter & direction    & (meV/3ML) \\
\hline
$P4/nmm$ & ---              & ---          & $-$639.2 \\
$C2/m$   & $E_g(\eta,\eta)$ & [110]        & $-$739.0 \\
$Abm2$   & $E_u(\eta,\eta)$ & [110]        & $-$760.1 \\
$P2_1/m$ & $E_g(\eta,0)$    & [100]        & $-$933.1 \\
$Pmn2_1$ & $E_u(\eta,0)$    & [100]        & $-$971.9 \\
\hline
$Pmmn$   & ---              & ---          & $-$905.8 \\
$P2_1/m$ & $B_{3g}$         & [100]        & $-$908.9 \\
$Pmn2_1$ & $B_{3u}$         & [100]        & $-$924.7 \\
$P2_1/m$ & $B_{2g}$         & [010]        & $-$933.2 \\
$Pmn2_1$ & $B_{2u}$         & [010]        & {\bf $-$971.9} \\
\end{tabular}
\end{ruledtabular}
\end{table}

\begin{figure}
\includegraphics{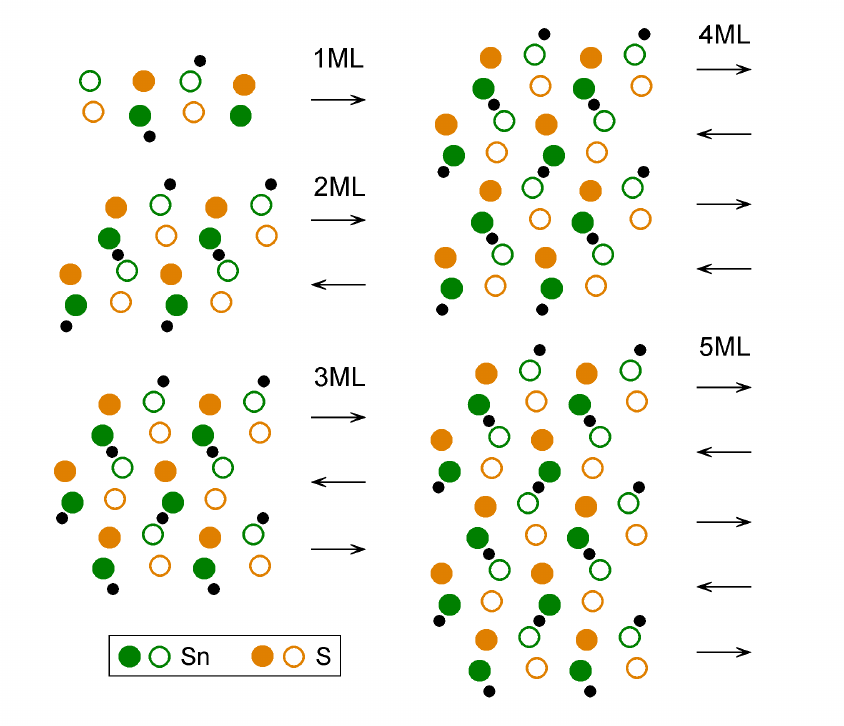}
\caption{Schematic drawing of distortions in ground-state structures of
nanoplatelets with different thickness in the $(0yz)$ plane. The filled symbols
denote atoms located at $x = 0$ and the closed symbols denote atoms located at
$x = \pm 1/2$. The arrows show the direction of polarization in monolayers.
The positions of Sn $s^2$ lone pairs are shown by black dots.}
\label{fig4}
\end{figure}

Surprisingly, the search for the ground-state structure of three-layer (3ML)
SnS nanoplatelets reveals that they become ferroelectric again. The phonon
spectrum of the $Pmmn$ phase exhibits four nondegenerate unstable modes at the
$\Gamma$~point, whereas the phonon spectrum of the $P4/nmm$ phase exhibits four
doubly degenerate unstable modes (two ferroelectric $E_u$ and two antiferroelectric
$E_g$ modes). As the eigenvectors of the unstable phonon modes become very
complicated, the comparison of the energies of the different low-symmetry phases
turns out to be more informative. The comparison of the energies of nanoplatelets
conjugated in different ways (Table~\ref{table3})
shows that the $Pmmn$ parent structure obtained using the $Cmcm$ conjugation
scheme is energetically more favorable as compared to the $P4/nmm$ parent
structure. The ground-state structure of the 3ML nanoplatelet is the polar
$Pmn2_1$ phase polarized along the armchair direction (see also
Fig.~\ref{fig4}).

\begin{table}
\caption{\label{table4}Energies of different phases of 4ML SnS nanoplatelets
obtained from the $P4/nmm$ and $Pmma$ parent structures. The energy of four
isolated monolayers is taken as the energy origin. The energy of the most
stable phase is in boldface.}
\begin{ruledtabular}
\begin{tabular}{cccc}
Phase    & Unstable phonon     & Displacement & Energy \\
         & and order parameter & direction    & (meV/4ML) \\
\hline
$P4/nmm$ & ---                 & ---          & $-$1005.9 \\
$Abm2$   & $E_u(\eta,\eta)$    & [110]        & $-$1096.3 \\
$C2/m$   & $E_g(\eta,\eta)$    & [110]        & $-$1148.6 \\
$Pmn2_1$ & $E_u(\eta,0)$       & [100]        & $-$1291.7 \\
$P2_1/m$ & $E_g(\eta,0)$       & [100]        & $-$1458.8 \\
\hline
$Pmma$   & ---                 & ---          & $-$1354.5 \\
$P2/m$   & $B_{3g}$            & [100]        & $-$1365.0 \\
$Pmc2_1$ & $B_{3u}$            & [100]        & $-$1382.7 \\
$P2_1/m$ & $B_{2g}$            & [010]        & $-$1383.1 \\
$Pma2$   & $B_{2u}$            & [010]        & $-$1399.9 \\
$P2_1/m$ & $B_{2g}$            & [010]        & {\bf $-$1459.9} \\
\end{tabular}
\end{ruledtabular}
\end{table}

The study of four-layer (4ML) SnS nanoplatelets reveals that the $Pmma$ phase
obtained using the $Cmcm$ conjugation scheme is the lowest-energy parent structure
(Table~\ref{table4}). The phonon spectrum of this phase exhibits six nondegenerate
unstable modes at the $\Gamma$~point, whereas the phonon spectrum of the $P4/nmm$
phase exhibits four doubly degenerate unstable modes at the $\Gamma$~point
(ferroelectric $E_u$ and antiferroelectric $E_g$ modes). It should be noted
that phonon spectra of this and thicker nanoplatelets exhibit several modes with
same symmetry, but with different displacement patterns. In this case, the
structural optimization usually gives the energy of the lowest-energy phase and
the calculation of the ``excited states'' needs additional cumbersome
orthogonalization procedure. This is why only the data for these low-energy
phases are presented in Tables~\ref{table4} and \ref{table5}. The ground-state
structure of the 4ML SnS nanoplatelet (the $P2_1/m$ phase) is not ferroelectric.

The study of five-layer (5ML) SnS nanoplatelets reveals that phonon spectra of
parent $P4/nmm$ and $Pnnm$ structures both exhibit eight unstable modes. The
energies of these parent phases and few lowest-energy structures for this
nanoplatelet are given in Table~\ref{table5}. It is seen that the lowest-energy
structure is the polar $Pmn2_1$ phase polarized along the armchair direction.
So, the nanoplatelets become ferroelectric again.

\begin{table}
\caption{\label{table5}Energies of different phases of 5ML SnS nanoplatelets
obtained from the $P4/nmm$ and $Pmmn$ parent structures. The energy of five
isolated monolayers is taken as the energy origin. The energy of the most stable
phase is in boldface.}
\begin{ruledtabular}
\begin{tabular}{cccc}
Phase & Unstable phonon     & Displacement & Energy \\
      & and order parameter & direction    & (meV/5ML) \\
\hline
$P4/nmm$    & ---              & ---          & $-$1384.3 \\
$Abm2$      & $E_u(\eta,\eta)$ & [110]        & $-$1500.0 \\
$C2/m$      & $E_g(\eta,\eta)$ & [110]        & $-$1519.0 \\
$P2_1/m$    & $E_g(\eta,0)$    & [100]        & $-$1903.0 \\
$Pmn2_1$    & $E_u(\eta,0)$    & [100]        & $-$1946.8 \\
\hline
$Pmmn$      & ---              & ---          & $-$1801.8 \\
$P2_1/m$    & $B_{3g}$         & [100]        & $-$1820.7 \\
$Pmn2_1$    & $B_{3u}$         & [100]        & $-$1839.3 \\
$P2_1/m$    & $B_{2g}$         & [010]        & $-$1903.1 \\
$Pmn2_1$    & $B_{2u}$         & [010]        & {\bf $-$1946.8} \\
\end{tabular}
\end{ruledtabular}
\end{table}

The explanation of such an ``intermittent'' ferroelectric properties of
nanoplatelets, in which the ferroelectricity appears only in nanoobjects with
odd number of monolayers, is quite simple. As mentioned above, the $Pnma$
structure of bulk SnS is actually antiferroelectric.~\cite{JPhysColl.29.C4-145}
When constructing a $n$ML nanoplatelet, its structure reproduces the structure
of bulk SnS (the energy of the corresponding conjunction is always lower). So,
when $n$ is even, the net polarization in the $xy$ plane is zero, but when $n$
is odd, the polarizations in the layers are not compensated and the net
polarization arises (Fig.~\ref{fig4}).

\section{Discussion}

Although our conclusion about the ground-state structure of SnS monolayers
(space group $Pmn2_1$) is consistent with earlier published
data,~\cite{ApplPhysLett.105.042103,JApplPhys.121.034302}
the small energy gain accompanying the formation of the polar phase
($\Delta U = 1.32$~meV per unit cell, Table~\ref{table6}) means that the phase
transition temperature in them is too low for practical purposes. This energy
is significantly less than the 38.3~meV value obtained in
Ref.~\onlinecite{PhysRevLett.117.097601}. In our opinion, such a strong
difference results from different exchange-correlation energy functionals used
in Ref.~\onlinecite{PhysRevLett.117.097601} (GGA-PBE) and in our calculations
(LDA). A~significant overestimation of the
lattice parameters in GGA calculations results in overestimated ferroelectric
instability and other related properties, including the energy gain and the
Curie temperature (1200~K for SnS monolayers according to
Ref.~\onlinecite{PhysRevLett.117.097601}).

\begin{table}
\caption{\label{table6}Physical properties of polar SnS monolayers (1ML) and
3ML and 5ML nanoplatelets. The specific elastic moduli are given in N/m, the
components of the specific piezoelectric $e$~tensor are in $10^{-10}$~C/m,
the components of the specific piezoelectric $d$~tensor are in pC/N, and the
specific polarization is in $10^{-10}$~C/m. The energy gain $\Delta U$ from
the ferroelectric ordering per one nanostructure is in meV.}
\begin{ruledtabular}
\begin{tabular}{ccccc}
Parameter & \multicolumn{2}{c}{1ML} & 3ML & 5ML \\
\cline{2-3}
          & This work & Literature data & & \\
\hline
$C_{xxxx}$ &  25.44 & 14.91\footnotemark[1]    & 77.1  & 124.2 \\
$C_{yyyy}$ &  42.33 & 35.97\footnotemark[1]    & 148.1 & 245.6 \\
$C_{xxyy}$ &  17.31 & 15.22\footnotemark[1]    & 62.2  & 101.2 \\
$C_{xyxy}$ &  23.69 & ---                      & 79.6  & 131.2 \\
$e_{xxx}$  &  42.4  & 18.1\footnotemark[1]     & 25.7  & 24.3 \\
$e_{xyy}$  &  4.13  & 13.8\footnotemark[1]     & 3.54  & 3.03 \\
$e_{yxy}$  &  50.2  & ---                      & 19.4  & 17.5 \\
$d_{xxx}$  &  221.5 & 144.8\footnotemark[1]    & 47.5  & 27.9 \\
$d_{xyy}$  & $-$80.8 & $-$22.9\footnotemark[1] & $-$17.6 & $-$10.3 \\
$d_{yxy}$  &  211.7 & ---                      & 24.3  & 13.3 \\
$P_s$      &  1.417 & 2.62\footnotemark[2], 2.47\footnotemark[3] & 2.15  & 2.35 \\
$\Delta U$ &  1.32  & 38.3\footnotemark[2]     & 66.2  & 145 \\
\end{tabular}
\end{ruledtabular}
\footnotetext[1]{Ref.~\onlinecite{ApplPhysLett.107.173104}.}
\footnotetext[2]{Ref.~\onlinecite{PhysRevLett.117.097601}.}
\footnotetext[3]{Ref.~\onlinecite{NanoLett.16.3236}.}
\end{table}

We now calculate some physical properties of SnS polar monolayers and 3ML and
5ML polar nanoplatelets. To characterize the properties of the nanoobjects,
the contribution of one nanoobject into these properties is usually computed.
These \emph{specific} properties, which were obtained by multiplying the
macroscopic property calculated for a supercell by the thickness occupied in
it by one nanoobject, are given in Table~\ref{table6}. In all cases, the
piezoelectric constant for fully relaxed geometry is reported as it is the only
parameter that can be compared with experiment (we can realize the situation
with clamped ions by performing measurements at optical frequencies, but
we are not able to change stress applied to a crystal so quickly).

For the $Pmn2_1$ structure of the monolayer (1ML), there are four non-zero
independent components of the specific elastic tensor and three non-zero
independent components of the specific piezoelectric tensor. The obtained
parameters are consistent with the results of earlier
calculations,~\cite{ApplPhysLett.107.173104,PhysRevLett.117.097601}
although sometimes they differ by a factor of two. These discrepancies
result mainly from different exchange-correlation functionals used in the
calculations. The new results in the table
are the elastic and piezoelectric tensor's components for the shear strain.
It is interesting that the $d_{yxy}$ piezoelectric coefficient for the shear
strain is as high as the $d_{xxx}$ coefficient.

When increasing the nanoplatelet thickness to 3ML and 5ML, the spontaneous
polarization $P_s$ increases, whereas the piezoelectric parameters, especially
those of the $d$~tensor, strongly decrease. The latter tendency is due to the
fact that the specific elastic moduli $C$ for nanoplatelets are defined per
nanostructure and change roughly proportional to its thickness. As a result,
the nanoplatelets become stiffer and their piezoelectric response decreases.

It should be noted that the energy gain of $\Delta U = 1.32$~meV per unit cell
for 1ML may be too small to produce the ferroelectric distortion in it. In
Ref.~\onlinecite{PhysSolidState.51.802}, a simple criterion was proposed
to estimate if quantum fluctuations can destroy the phase transition.
According to this criterion, in the case of 1ML the energy gain should be
$\ge 1.89$~meV to make the ferroelectric distortion stable against quantum
fluctuations.

A particular interest to polar SnS monolayers and nanoplatelets is due to the
fact that they present a new mechanism of ferroelectricity. The appearance of
ferroelectricity in these objects is not a consequence of the long-range
dipole-dipole interaction, but is a result of short-range interaction of lone
pairs of Sn$^{2+}$ ions (Fig.~\ref{fig4}).

In bulk SnS, the antiferroelectricity is attributed to the ordering of $s^2$
lone pairs of Sn$^{2+}$ ions whose interaction is strong (as follows from high
phase transition temperature in it) and short-range. In Sec.~\ref{sec3c}, it was
shown that the ferroelectricity in nanoplatelets is associated with uncompensated
polarization of monolayers. Strong short-range interaction of lone pairs
suggests that the distorted structure of nanoplatelets can be stable up to
high temperatures. In thick nanoplatelets, the Curie temperature can approach
that of the structural phase transition in SnS (878~K). For SnS monolayer,
the energy gain $\Delta U$ from the ferroelectric ordering (the energy barrier
separating two polar states) is too small, but it increases dramatically with
increasing number of monolayers (Table~\ref{table6}). Our calculations suggest
that the ferroelectric phase can be observed at 300~K in the nanoplatelets with
a thickness of 3~ML or 5~ML. In thicker nanoplatelets, the energy barrier becomes
too high to switch or rotate the polarization in them by an external electric
field. The polarization in nanoobjects increases with increasing their thickness
and approaches the local polarization in the layers of bulk SnS
($\sim 3.7 \cdot 10^{-10}$~C/m).

In addition to thermoelectric and piezoelectric applications of SnS monolayers
and nanoplatelets, one can propose yet another possible use of their polar
properties in electrically-driven thermal valves. Recent calculations have
predicted high anisotropy of phonon conductivity in monolayer
SnS.~\cite{JPhysChemC.120.18841} As the orientation of polarization in
nanoobjects can be switched by an electric field, a device whose thermal
conductivity is controlled by an electric field can be proposed.
These devices can be realized as the solid-state ones in which the nanoobjects
are embedded in a polymer matrix or as dense colloid solutions.

\section{Conclusions}

In this work, the ground-state structure of SnS monolayers and SnS nanoplatelets
with a thickness from two to five monolayers has been calculated from first
principles. All nanoobjects containing odd number of monolayers were shown to
be ferroelectric. A new way of creating switchable spontaneous polarization by
nanostructuring of an antiferroelectric compound whose distorted structure
is associated with the ordering of lone pairs was revealed. When this short-range
interaction is strong, one can anticipate high Curie temperature in these
ferroelectrics. The ferroelectric, piezoelectric, and elastic
properties of SnS polar nanoobjects were calculated. Based on the results
of this study, one can anticipate that similar phenomena can be observed
in other IV-VI layered semiconductors---GeS, GeSe, and SnSe. It would be also
interesting to study if the proposed mechanism of ferroelectricity is possible
in nanoplatelets made of other antiferroelectric compounds.

\section*{Supplementary material}

See supplementary material for the ground-state structures of bulk SnS and
considered nanoplatelets.

\begin{acknowledgments}
This work was supported by the Russian Foundation for Basic Research (Grant
No.~17-02-01068).
\end{acknowledgments}


\begin{thebibliography}{33}%
\makeatletter
\providecommand \@ifxundefined [1]{%
 \@ifx{#1\undefined}
}%
\providecommand \@ifnum [1]{%
 \ifnum #1\expandafter \@firstoftwo
 \else \expandafter \@secondoftwo
 \fi
}%
\providecommand \@ifx [1]{%
 \ifx #1\expandafter \@firstoftwo
 \else \expandafter \@secondoftwo
 \fi
}%
\providecommand \natexlab [1]{#1}%
\providecommand \enquote  [1]{``#1''}%
\providecommand \bibnamefont  [1]{#1}%
\providecommand \bibfnamefont [1]{#1}%
\providecommand \citenamefont [1]{#1}%
\providecommand \href@noop [0]{\@secondoftwo}%
\providecommand \href [0]{\begingroup \@sanitize@url \@href}%
\providecommand \@href[1]{\@@startlink{#1}\@@href}%
\providecommand \@@href[1]{\endgroup#1\@@endlink}%
\providecommand \@sanitize@url [0]{\catcode `\\12\catcode `\$12\catcode
  `\&12\catcode `\#12\catcode `\^12\catcode `\_12\catcode `\%12\relax}%
\providecommand \@@startlink[1]{}%
\providecommand \@@endlink[0]{}%
\providecommand \url  [0]{\begingroup\@sanitize@url \@url }%
\providecommand \@url [1]{\endgroup\@href {#1}{\urlprefix }}%
\providecommand \urlprefix  [0]{URL }%
\providecommand \Eprint [0]{\href }%
\providecommand \doibase [0]{http://dx.doi.org/}%
\providecommand \selectlanguage [0]{\@gobble}%
\providecommand \bibinfo  [0]{\@secondoftwo}%
\providecommand \bibfield  [0]{\@secondoftwo}%
\providecommand \translation [1]{[#1]}%
\providecommand \BibitemOpen [0]{}%
\providecommand \bibitemStop [0]{}%
\providecommand \bibitemNoStop [0]{.\EOS\space}%
\providecommand \EOS [0]{\spacefactor3000\relax}%
\providecommand \BibitemShut  [1]{\csname bibitem#1\endcsname}%
\let\auto@bib@innerbib\@empty
\bibitem [{\citenamefont {Ithurria}\ and\ \citenamefont
  {Dubertret}(2008)}]{JAmChemSoc.130.16504}%
  \BibitemOpen
  \bibfield  {author} {\bibinfo {author} {\bibfnamefont {S.}~\bibnamefont
  {Ithurria}}\ and\ \bibinfo {author} {\bibfnamefont {B.}~\bibnamefont
  {Dubertret}},\ }\href {\doibase 10.1021/ja807724e} {\bibfield  {journal}
  {\bibinfo  {journal} {J. Am. Chem. Soc.}\ }\textbf {\bibinfo {volume}
  {130}},\ \bibinfo {pages} {16504} (\bibinfo {year} {2008})}\BibitemShut
  {NoStop}%
\bibitem [{\citenamefont {Osada}\ and\ \citenamefont
  {Sasaki}(2009)}]{JMaterChem.19.2503}%
  \BibitemOpen
  \bibfield  {author} {\bibinfo {author} {\bibfnamefont {M.}~\bibnamefont
  {Osada}}\ and\ \bibinfo {author} {\bibfnamefont {T.}~\bibnamefont {Sasaki}},\
  }\href {\doibase 10.1039/B820160A} {\bibfield  {journal} {\bibinfo  {journal}
  {J. Mater. Chem.}\ }\textbf {\bibinfo {volume} {19}},\ \bibinfo {pages}
  {2503} (\bibinfo {year} {2009})}\BibitemShut {NoStop}%
\bibitem [{\citenamefont {Choi}\ \emph {et~al.}(2010)\citenamefont {Choi},
  \citenamefont {Lahiri}, \citenamefont {Seelaboyina},\ and\ \citenamefont
  {Kang}}]{CritRevSolidStateMaterSci.35.52}%
  \BibitemOpen
  \bibfield  {author} {\bibinfo {author} {\bibfnamefont {W.}~\bibnamefont
  {Choi}}, \bibinfo {author} {\bibfnamefont {I.}~\bibnamefont {Lahiri}},
  \bibinfo {author} {\bibfnamefont {R.}~\bibnamefont {Seelaboyina}}, \ and\
  \bibinfo {author} {\bibfnamefont {Y.~S.}\ \bibnamefont {Kang}},\ }\href
  {\doibase 10.1080/10408430903505036} {\bibfield  {journal} {\bibinfo
  {journal} {Crit. Rev. Solid State Mater. Sci.}\ }\textbf {\bibinfo {volume}
  {35}},\ \bibinfo {pages} {52} (\bibinfo {year} {2010})}\BibitemShut {NoStop}%
\bibitem [{\citenamefont {Huang}, \citenamefont {Zeng},\ and\ \citenamefont
  {Zhang}(2013)}]{ChemSocRev.42.1934}%
  \BibitemOpen
  \bibfield  {author} {\bibinfo {author} {\bibfnamefont {X.}~\bibnamefont
  {Huang}}, \bibinfo {author} {\bibfnamefont {Z.}~\bibnamefont {Zeng}}, \ and\
  \bibinfo {author} {\bibfnamefont {H.}~\bibnamefont {Zhang}},\ }\href
  {\doibase 10.1039/C2CS35387C} {\bibfield  {journal} {\bibinfo  {journal}
  {Chem. Soc. Rev.}\ }\textbf {\bibinfo {volume} {42}},\ \bibinfo {pages}
  {1934} (\bibinfo {year} {2013})}\BibitemShut {NoStop}%
\bibitem [{\citenamefont {Biacchi}, \citenamefont {\mbox{Vaughn II}},\ and\
  \citenamefont {Schaak}(2013)}]{JAmChemSoc.135.11634}%
  \BibitemOpen
  \bibfield  {author} {\bibinfo {author} {\bibfnamefont {A.~J.}\ \bibnamefont
  {Biacchi}}, \bibinfo {author} {\bibfnamefont {D.~D.}\ \bibnamefont
  {\mbox{Vaughn II}}}, \ and\ \bibinfo {author} {\bibfnamefont {R.~E.}\
  \bibnamefont {Schaak}},\ }\href {\doibase 10.1021/ja405203e} {\bibfield
  {journal} {\bibinfo  {journal} {J. Am. Chem. Soc.}\ }\textbf {\bibinfo
  {volume} {135}},\ \bibinfo {pages} {11634} (\bibinfo {year}
  {2013})}\BibitemShut {NoStop}%
\bibitem [{\citenamefont {de~Kergommeaux}\ \emph {et~al.}(2015)\citenamefont
  {de~Kergommeaux}, \citenamefont {Lopez-Haro}, \citenamefont {Pouget},
  \citenamefont {Zuo}, \citenamefont {Lebrun}, \citenamefont {Chandezon},
  \citenamefont {Aldakov},\ and\ \citenamefont {Reiss}}]{JAmChemSoc.137.9943}%
  \BibitemOpen
  \bibfield  {author} {\bibinfo {author} {\bibfnamefont {A.}~\bibnamefont
  {de~Kergommeaux}}, \bibinfo {author} {\bibfnamefont {M.}~\bibnamefont
  {Lopez-Haro}}, \bibinfo {author} {\bibfnamefont {S.}~\bibnamefont {Pouget}},
  \bibinfo {author} {\bibfnamefont {J.-M.}\ \bibnamefont {Zuo}}, \bibinfo
  {author} {\bibfnamefont {C.}~\bibnamefont {Lebrun}}, \bibinfo {author}
  {\bibfnamefont {F.}~\bibnamefont {Chandezon}}, \bibinfo {author}
  {\bibfnamefont {D.}~\bibnamefont {Aldakov}}, \ and\ \bibinfo {author}
  {\bibfnamefont {P.}~\bibnamefont {Reiss}},\ }\href {\doibase
  10.1021/jacs.5b05576} {\bibfield  {journal} {\bibinfo  {journal} {J. Am.
  Chem. Soc.}\ }\textbf {\bibinfo {volume} {137}},\ \bibinfo {pages} {9943}
  (\bibinfo {year} {2015})}\BibitemShut {NoStop}%
\bibitem [{\citenamefont {Brent}\ \emph {et~al.}(2015)\citenamefont {Brent},
  \citenamefont {Lewis}, \citenamefont {Lorenz}, \citenamefont {Lewis},
  \citenamefont {Savjani}, \citenamefont {Haigh}, \citenamefont {Seifert},
  \citenamefont {Derby},\ and\ \citenamefont {O'Brien}}]{JAmChemSoc.137.12689}%
  \BibitemOpen
  \bibfield  {author} {\bibinfo {author} {\bibfnamefont {J.~R.}\ \bibnamefont
  {Brent}}, \bibinfo {author} {\bibfnamefont {D.~J.}\ \bibnamefont {Lewis}},
  \bibinfo {author} {\bibfnamefont {T.}~\bibnamefont {Lorenz}}, \bibinfo
  {author} {\bibfnamefont {E.~A.}\ \bibnamefont {Lewis}}, \bibinfo {author}
  {\bibfnamefont {N.}~\bibnamefont {Savjani}}, \bibinfo {author} {\bibfnamefont
  {S.~J.}\ \bibnamefont {Haigh}}, \bibinfo {author} {\bibfnamefont
  {G.}~\bibnamefont {Seifert}}, \bibinfo {author} {\bibfnamefont
  {B.}~\bibnamefont {Derby}}, \ and\ \bibinfo {author} {\bibfnamefont
  {P.}~\bibnamefont {O'Brien}},\ }\href {\doibase 10.1021/jacs.5b08236}
  {\bibfield  {journal} {\bibinfo  {journal} {J. Am. Chem. Soc.}\ }\textbf
  {\bibinfo {volume} {137}},\ \bibinfo {pages} {12689} (\bibinfo {year}
  {2015})}\BibitemShut {NoStop}%
\bibitem [{\citenamefont {Tritsaris}, \citenamefont {Malone},\ and\
  \citenamefont {Kaxiras}(2013)}]{JApplPhys.113.233507}%
  \BibitemOpen
  \bibfield  {author} {\bibinfo {author} {\bibfnamefont {G.~A.}\ \bibnamefont
  {Tritsaris}}, \bibinfo {author} {\bibfnamefont {B.~D.}\ \bibnamefont
  {Malone}}, \ and\ \bibinfo {author} {\bibfnamefont {E.}~\bibnamefont
  {Kaxiras}},\ }\href {\doibase 10.1063/1.4811455} {\bibfield  {journal}
  {\bibinfo  {journal} {J. Appl. Phys.}\ }\textbf {\bibinfo {volume} {113}},\
  \bibinfo {pages} {233507} (\bibinfo {year} {2013})}\BibitemShut {NoStop}%
\bibitem [{\citenamefont {Singh}\ and\ \citenamefont
  {Hennig}(2014)}]{ApplPhysLett.105.042103}%
  \BibitemOpen
  \bibfield  {author} {\bibinfo {author} {\bibfnamefont {A.~K.}\ \bibnamefont
  {Singh}}\ and\ \bibinfo {author} {\bibfnamefont {R.~G.}\ \bibnamefont
  {Hennig}},\ }\href {\doibase 10.1063/1.4891230} {\bibfield  {journal}
  {\bibinfo  {journal} {Appl. Phys. Lett.}\ }\textbf {\bibinfo {volume}
  {105}},\ \bibinfo {pages} {042103} (\bibinfo {year} {2014})}\BibitemShut
  {NoStop}%
\bibitem [{\citenamefont {Gomes}\ and\ \citenamefont
  {Carvalho}(2015)}]{PhysRevB.92.085406}%
  \BibitemOpen
  \bibfield  {author} {\bibinfo {author} {\bibfnamefont {L.~C.}\ \bibnamefont
  {Gomes}}\ and\ \bibinfo {author} {\bibfnamefont {A.}~\bibnamefont
  {Carvalho}},\ }\href {\doibase 10.1103/PhysRevB.92.085406} {\bibfield
  {journal} {\bibinfo  {journal} {Phys. Rev. B}\ }\textbf {\bibinfo {volume}
  {92}},\ \bibinfo {pages} {085406} (\bibinfo {year} {2015})}\BibitemShut
  {NoStop}%
\bibitem [{\citenamefont {Huang}, \citenamefont {Wu},\ and\ \citenamefont
  {Li}(2016)}]{JChemPhys.144.114708}%
  \BibitemOpen
  \bibfield  {author} {\bibinfo {author} {\bibfnamefont {L.}~\bibnamefont
  {Huang}}, \bibinfo {author} {\bibfnamefont {F.}~\bibnamefont {Wu}}, \ and\
  \bibinfo {author} {\bibfnamefont {J.}~\bibnamefont {Li}},\ }\href {\doibase
  10.1063/1.4943969} {\bibfield  {journal} {\bibinfo  {journal} {J. Chem.
  Phys.}\ }\textbf {\bibinfo {volume} {144}},\ \bibinfo {pages} {114708}
  (\bibinfo {year} {2016})}\BibitemShut {NoStop}%
\bibitem [{\citenamefont {Sandonas}\ \emph {et~al.}(2016)\citenamefont
  {Sandonas}, \citenamefont {Teich}, \citenamefont {Gutierrez}, \citenamefont
  {Lorenz}, \citenamefont {Pecchia}, \citenamefont {Seifert},\ and\
  \citenamefont {Cuniberti}}]{JPhysChemC.120.18841}%
  \BibitemOpen
  \bibfield  {author} {\bibinfo {author} {\bibfnamefont {L.~M.}\ \bibnamefont
  {Sandonas}}, \bibinfo {author} {\bibfnamefont {D.}~\bibnamefont {Teich}},
  \bibinfo {author} {\bibfnamefont {R.}~\bibnamefont {Gutierrez}}, \bibinfo
  {author} {\bibfnamefont {T.}~\bibnamefont {Lorenz}}, \bibinfo {author}
  {\bibfnamefont {A.}~\bibnamefont {Pecchia}}, \bibinfo {author} {\bibfnamefont
  {G.}~\bibnamefont {Seifert}}, \ and\ \bibinfo {author} {\bibfnamefont
  {G.}~\bibnamefont {Cuniberti}},\ }\href {\doibase 10.1021/acs.jpcc.6b04969}
  {\bibfield  {journal} {\bibinfo  {journal} {J. Phys. Chem. C}\ }\textbf
  {\bibinfo {volume} {120}},\ \bibinfo {pages} {18841} (\bibinfo {year}
  {2016})}\BibitemShut {NoStop}%
\bibitem [{\citenamefont {Wu}\ and\ \citenamefont
  {Zeng}(2016)}]{NanoLett.16.3236}%
  \BibitemOpen
  \bibfield  {author} {\bibinfo {author} {\bibfnamefont {M.}~\bibnamefont
  {Wu}}\ and\ \bibinfo {author} {\bibfnamefont {X.~C.}\ \bibnamefont {Zeng}},\
  }\href {\doibase 10.1021/acs.nanolett.6b00726} {\bibfield  {journal}
  {\bibinfo  {journal} {Nano Lett.}\ }\textbf {\bibinfo {volume} {16}},\
  \bibinfo {pages} {3236} (\bibinfo {year} {2016})}\BibitemShut {NoStop}%
\bibitem [{\citenamefont {Guo}\ and\ \citenamefont
  {Wang}(2017)}]{JApplPhys.121.034302}%
  \BibitemOpen
  \bibfield  {author} {\bibinfo {author} {\bibfnamefont {S.-D.}\ \bibnamefont
  {Guo}}\ and\ \bibinfo {author} {\bibfnamefont {Y.-H.}\ \bibnamefont {Wang}},\
  }\href {\doibase 10.1063/1.4974200} {\bibfield  {journal} {\bibinfo
  {journal} {J. Appl. Phys.}\ }\textbf {\bibinfo {volume} {121}},\ \bibinfo
  {pages} {034302} (\bibinfo {year} {2017})}\BibitemShut {NoStop}%
\bibitem [{\citenamefont {Xu}\ \emph {et~al.}(2017)\citenamefont {Xu},
  \citenamefont {Yang}, \citenamefont {Wang},\ and\ \citenamefont
  {Feng}}]{PhysRevB.95.235434}%
  \BibitemOpen
  \bibfield  {author} {\bibinfo {author} {\bibfnamefont {L.}~\bibnamefont
  {Xu}}, \bibinfo {author} {\bibfnamefont {M.}~\bibnamefont {Yang}}, \bibinfo
  {author} {\bibfnamefont {S.~J.}\ \bibnamefont {Wang}}, \ and\ \bibinfo
  {author} {\bibfnamefont {Y.~P.}\ \bibnamefont {Feng}},\ }\href {\doibase
  10.1103/PhysRevB.95.235434} {\bibfield  {journal} {\bibinfo  {journal} {Phys.
  Rev. B}\ }\textbf {\bibinfo {volume} {95}},\ \bibinfo {pages} {235434}
  (\bibinfo {year} {2017})}\BibitemShut {NoStop}%
\bibitem [{\citenamefont {Tian}\ \emph {et~al.}(2017)\citenamefont {Tian},
  \citenamefont {Guo}, \citenamefont {Zhao}, \citenamefont {Li},\ and\
  \citenamefont {Xue}}]{ACSNano.11.2219}%
  \BibitemOpen
  \bibfield  {author} {\bibinfo {author} {\bibfnamefont {Z.}~\bibnamefont
  {Tian}}, \bibinfo {author} {\bibfnamefont {C.}~\bibnamefont {Guo}}, \bibinfo
  {author} {\bibfnamefont {M.}~\bibnamefont {Zhao}}, \bibinfo {author}
  {\bibfnamefont {R.}~\bibnamefont {Li}}, \ and\ \bibinfo {author}
  {\bibfnamefont {J.}~\bibnamefont {Xue}},\ }\href {\doibase
  10.1021/acsnano.6b08704} {\bibfield  {journal} {\bibinfo  {journal} {ACS
  Nano}\ }\textbf {\bibinfo {volume} {11}},\ \bibinfo {pages} {2219} (\bibinfo
  {year} {2017})}\BibitemShut {NoStop}%
\bibitem [{\citenamefont {Haleoot}\ \emph {et~al.}(2017)\citenamefont
  {Haleoot}, \citenamefont {Paillard}, \citenamefont {Kaloni}, \citenamefont
  {Mehboudi}, \citenamefont {Xu}, \citenamefont {Bellaiche},\ and\
  \citenamefont {Barraza-Lopez}}]{PhysRevLett.118.227401}%
  \BibitemOpen
  \bibfield  {author} {\bibinfo {author} {\bibfnamefont {R.}~\bibnamefont
  {Haleoot}}, \bibinfo {author} {\bibfnamefont {C.}~\bibnamefont {Paillard}},
  \bibinfo {author} {\bibfnamefont {T.~P.}\ \bibnamefont {Kaloni}}, \bibinfo
  {author} {\bibfnamefont {M.}~\bibnamefont {Mehboudi}}, \bibinfo {author}
  {\bibfnamefont {B.}~\bibnamefont {Xu}}, \bibinfo {author} {\bibfnamefont
  {L.}~\bibnamefont {Bellaiche}}, \ and\ \bibinfo {author} {\bibfnamefont
  {S.}~\bibnamefont {Barraza-Lopez}},\ }\href {\doibase
  10.1103/PhysRevLett.118.227401} {\bibfield  {journal} {\bibinfo  {journal}
  {Phys. Rev. Lett.}\ }\textbf {\bibinfo {volume} {118}},\ \bibinfo {pages}
  {227401} (\bibinfo {year} {2017})}\BibitemShut {NoStop}%
\bibitem [{\citenamefont {Fei}\ \emph {et~al.}(2015)\citenamefont {Fei},
  \citenamefont {Li}, \citenamefont {Li},\ and\ \citenamefont
  {Yang}}]{ApplPhysLett.107.173104}%
  \BibitemOpen
  \bibfield  {author} {\bibinfo {author} {\bibfnamefont {R.}~\bibnamefont
  {Fei}}, \bibinfo {author} {\bibfnamefont {W.}~\bibnamefont {Li}}, \bibinfo
  {author} {\bibfnamefont {J.}~\bibnamefont {Li}}, \ and\ \bibinfo {author}
  {\bibfnamefont {L.}~\bibnamefont {Yang}},\ }\href {\doibase
  10.1063/1.4934750} {\bibfield  {journal} {\bibinfo  {journal} {Appl. Phys.
  Lett.}\ }\textbf {\bibinfo {volume} {107}},\ \bibinfo {pages} {173104}
  (\bibinfo {year} {2015})}\BibitemShut {NoStop}%
\bibitem [{\citenamefont {Hanakata}\ \emph {et~al.}(2016)\citenamefont
  {Hanakata}, \citenamefont {Carvalho}, \citenamefont {Campbell},\ and\
  \citenamefont {Park}}]{PhysRevB.94.035304}%
  \BibitemOpen
  \bibfield  {author} {\bibinfo {author} {\bibfnamefont {P.~Z.}\ \bibnamefont
  {Hanakata}}, \bibinfo {author} {\bibfnamefont {A.}~\bibnamefont {Carvalho}},
  \bibinfo {author} {\bibfnamefont {D.~K.}\ \bibnamefont {Campbell}}, \ and\
  \bibinfo {author} {\bibfnamefont {H.~S.}\ \bibnamefont {Park}},\ }\href
  {\doibase 10.1103/PhysRevB.94.035304} {\bibfield  {journal} {\bibinfo
  {journal} {Phys. Rev. B}\ }\textbf {\bibinfo {volume} {94}},\ \bibinfo
  {pages} {035304} (\bibinfo {year} {2016})}\BibitemShut {NoStop}%
\bibitem [{\citenamefont {Chattopadyay}, \citenamefont {Pannetier},\ and\
  \citenamefont {von Schnering}(1986)}]{JPhysChemSolids.47.879}%
  \BibitemOpen
  \bibfield  {author} {\bibinfo {author} {\bibfnamefont {T.}~\bibnamefont
  {Chattopadyay}}, \bibinfo {author} {\bibfnamefont {J.}~\bibnamefont
  {Pannetier}}, \ and\ \bibinfo {author} {\bibfnamefont {H.~G.}\ \bibnamefont
  {von Schnering}},\ }\href {\doibase 10.1016/0022-3697(86)90059-4} {\bibfield
  {journal} {\bibinfo  {journal} {J. Phys. Chem. Solids}\ }\textbf {\bibinfo
  {volume} {47}},\ \bibinfo {pages} {879} (\bibinfo {year} {1986})}\BibitemShut
  {NoStop}%
\bibitem [{\citenamefont {Mariano}\ and\ \citenamefont
  {Chopra}(1967)}]{ApplPhysLett.10.282}%
  \BibitemOpen
  \bibfield  {author} {\bibinfo {author} {\bibfnamefont {A.~N.}\ \bibnamefont
  {Mariano}}\ and\ \bibinfo {author} {\bibfnamefont {K.~L.}\ \bibnamefont
  {Chopra}},\ }\href {\doibase 10.1063/1.1754812} {\bibfield  {journal}
  {\bibinfo  {journal} {Appl. Phys. Lett.}\ }\textbf {\bibinfo {volume} {10}},\
  \bibinfo {pages} {282} (\bibinfo {year} {1967})}\BibitemShut {NoStop}%
\bibitem [{\citenamefont {Fei}, \citenamefont {Kang},\ and\ \citenamefont
  {Yang}(2016)}]{PhysRevLett.117.097601}%
  \BibitemOpen
  \bibfield  {author} {\bibinfo {author} {\bibfnamefont {R.}~\bibnamefont
  {Fei}}, \bibinfo {author} {\bibfnamefont {W.}~\bibnamefont {Kang}}, \ and\
  \bibinfo {author} {\bibfnamefont {L.}~\bibnamefont {Yang}},\ }\href {\doibase
  10.1103/PhysRevLett.117.097601} {\bibfield  {journal} {\bibinfo  {journal}
  {Phys. Rev. Lett.}\ }\textbf {\bibinfo {volume} {117}},\ \bibinfo {pages}
  {097601} (\bibinfo {year} {2016})}\BibitemShut {NoStop}%
\bibitem [{Note1()}]{Note1}%
  \BibitemOpen
  \bibinfo {note} {Our calculations using the GGA-PBE PAW pseudopotentials
  taken from Ref.~\protect \rev@citealpnum {ComputMaterSci.81.446} gave the
  energy gain of 20.4~meV per unit cell for SnS monolayer with the $Pmn2_1$
  structure. Using of the LDA PAW pseudopotentials \protect from the
  same source gave much lower energy, 0.48~meV per unit cell.}\BibitemShut
  {Stop}%
\bibitem [{\citenamefont {Marini}, \citenamefont {Garc{\'i}a-Gonz{\'a}lez},\
  and\ \citenamefont {Rubio}(2006)}]{PhysRevLett.96.136404}%
  \BibitemOpen
  \bibfield  {author} {\bibinfo {author} {\bibfnamefont {A.}~\bibnamefont
  {Marini}}, \bibinfo {author} {\bibfnamefont {P.}~\bibnamefont
  {Garc{\'i}a-Gonz{\'a}lez}}, \ and\ \bibinfo {author} {\bibfnamefont
  {A.}~\bibnamefont {Rubio}},\ }\href {\doibase 10.1103/PhysRevLett.96.136404}
  {\bibfield  {journal} {\bibinfo  {journal} {Phys. Rev. Lett.}\ }\textbf
  {\bibinfo {volume} {96}},\ \bibinfo {pages} {136404} (\bibinfo {year}
  {2006})}\BibitemShut {NoStop}%
\bibitem [{\citenamefont {Rappe}\ \emph {et~al.}(1990)\citenamefont {Rappe},
  \citenamefont {Rabe}, \citenamefont {Kaxiras},\ and\ \citenamefont
  {Joannopoulos}}]{PhysRevB.41.1227}%
  \BibitemOpen
  \bibfield  {author} {\bibinfo {author} {\bibfnamefont {A.~M.}\ \bibnamefont
  {Rappe}}, \bibinfo {author} {\bibfnamefont {K.~M.}\ \bibnamefont {Rabe}},
  \bibinfo {author} {\bibfnamefont {E.}~\bibnamefont {Kaxiras}}, \ and\
  \bibinfo {author} {\bibfnamefont {J.~D.}\ \bibnamefont {Joannopoulos}},\
  }\href {\doibase 10.1103/PhysRevB.41.1227} {\bibfield  {journal} {\bibinfo
  {journal} {Phys. Rev. B}\ }\textbf {\bibinfo {volume} {41}},\ \bibinfo
  {pages} {1227} (\bibinfo {year} {1990})}\BibitemShut {NoStop}%
\bibitem [{\citenamefont
  {Lebedev}(2009{\natexlab{a}})}]{PhysSolidState.51.362}%
  \BibitemOpen
  \bibfield  {author} {\bibinfo {author} {\bibfnamefont {A.~I.}\ \bibnamefont
  {Lebedev}},\ }\href {\doibase 10.1134/S1063783409020279} {\bibfield
  {journal} {\bibinfo  {journal} {Phys. Solid State}\ }\textbf {\bibinfo
  {volume} {51}},\ \bibinfo {pages} {362} (\bibinfo {year}
  {2009}{\natexlab{a}})}\BibitemShut {NoStop}%
\bibitem [{Note2()}]{Note2}%
  \BibitemOpen
  \bibinfo {note} {Large Born effective charge of Sn in the $Cmcm$ phase ($Z^*
  = {}$5.4--5.7 in the $xy$ plane) favors the appearance of ferroelectricity,
  but the strong distortion of the structure \protect associated with
  the stereochemical activity of the $s^2$ lone pair of Sn$^{2+}$ ions makes
  this compound antiferroelectric.}\BibitemShut {Stop}%
\bibitem [{Note3()}]{Note3}%
  \BibitemOpen
  \bibinfo {note} {In order to get better accuracy for phonon frequencies on
  the boundary of the Brillouin zone, the response function calculations
  \protect were also performed on supercells containing two
  monolayers in the unit cell. In this case, the $Z$~point of the Brillouin
  zone is projected on the $\Gamma $~point.}\BibitemShut {Stop}%
\bibitem [{\citenamefont {Dion}\ \emph {et~al.}(2004)\citenamefont {Dion},
  \citenamefont {Rydberg}, \citenamefont {Schr{\"o}der}, \citenamefont
  {Langreth},\ and\ \citenamefont {Lundqvist}}]{PhysRevLett.92.246401}%
  \BibitemOpen
  \bibfield  {author} {\bibinfo {author} {\bibfnamefont {M.}~\bibnamefont
  {Dion}}, \bibinfo {author} {\bibfnamefont {H.}~\bibnamefont {Rydberg}},
  \bibinfo {author} {\bibfnamefont {E.}~\bibnamefont {Schr{\"o}der}}, \bibinfo
  {author} {\bibfnamefont {D.~C.}\ \bibnamefont {Langreth}}, \ and\ \bibinfo
  {author} {\bibfnamefont {B.~I.}\ \bibnamefont {Lundqvist}},\ }\href {\doibase
  10.1103/PhysRevLett.92.246401} {\bibfield  {journal} {\bibinfo  {journal}
  {Phys. Rev. Lett.}\ }\textbf {\bibinfo {volume} {92}},\ \bibinfo {pages}
  {246401} (\bibinfo {year} {2004})}\BibitemShut {NoStop}%
\bibitem [{Note4()}]{Note4}%
  \BibitemOpen
  \bibinfo {note} {\protect The obtained results are independent of
  the exchange-correlation functional used in the calculations. The same
  sequence of the energies was obtained when modeling different low-symmetry
  phases of 2ML nanoplatelets using the GGA-PBE and dispersion-corrected
  GGA-PBE+D2 functionals.}\BibitemShut {Stop}%
\bibitem [{\citenamefont {Pawley}(1968)}]{JPhysColl.29.C4-145}%
  \BibitemOpen
  \bibfield  {author} {\bibinfo {author} {\bibfnamefont {G.~S.}\ \bibnamefont
  {Pawley}},\ }\href {\doibase 10.1051/jphyscol:1968423} {\bibfield  {journal}
  {\bibinfo  {journal} {J. Phys. Colloq.}\ }\textbf {\bibinfo {volume}
  {29(C4)}},\ \bibinfo {pages} {145} (\bibinfo {year} {1968})}\BibitemShut
  {NoStop}%
\bibitem [{\citenamefont
  {Lebedev}(2009{\natexlab{b}})}]{PhysSolidState.51.802}%
  \BibitemOpen
  \bibfield  {author} {\bibinfo {author} {\bibfnamefont {A.~I.}\ \bibnamefont
  {Lebedev}},\ }\href {\doibase 10.1134/S1063783409040283} {\bibfield
  {journal} {\bibinfo  {journal} {Phys. Solid State}\ }\textbf {\bibinfo
  {volume} {51}},\ \bibinfo {pages} {802} (\bibinfo {year}
  {2009}{\natexlab{b}})}\BibitemShut {NoStop}%
\bibitem [{\citenamefont {Garrity}\ \emph {et~al.}(2014)\citenamefont
  {Garrity}, \citenamefont {Bennett}, \citenamefont {Rabe},\ and\ \citenamefont
  {Vanderbilt}}]{ComputMaterSci.81.446}%
  \BibitemOpen
  \bibfield  {author} {\bibinfo {author} {\bibfnamefont {K.~F.}\ \bibnamefont
  {Garrity}}, \bibinfo {author} {\bibfnamefont {J.~W.}\ \bibnamefont
  {Bennett}}, \bibinfo {author} {\bibfnamefont {K.~M.}\ \bibnamefont {Rabe}}, \
  and\ \bibinfo {author} {\bibfnamefont {D.}~\bibnamefont {Vanderbilt}},\
  }\href {\doibase 10.1016/j.commatsci.2013.08.053} {\bibfield  {journal}
  {\bibinfo  {journal} {Comput. Mater. Sci.}\ }\textbf {\bibinfo {volume}
  {81}},\ \bibinfo {pages} {446} (\bibinfo {year} {2014})}\BibitemShut
  {NoStop}%
\end{thebibliography}

\providecommand{\BIBYu}{Yu}

\end{document}